\documentstyle[pra,aps,preprint,epsf]{revtex}

\def\Namol{\mbox{Na}_2}
\def\Xstate{\mbox{X}\,{}^1\Sigma_g^+}
\def\Astate{\mbox{A}\,{}^1\Sigma_u^+}
\def\Bstate{\mbox{B}\,{}^1\Pi_u}

\def\astate{\mbox{a}\,{}^3\Sigma_u^+}
\def\bstate{\mbox{b}\,{}^3\Sigma_g^+}
\def\cstate{\mbox{c}\,{}^3\Pi_g}

\def\invcm{\mbox{cm}^{-1}}

\begin{document}
\draft

\title{Theoretical study of the absorption spectra of the sodium dimer}
\author{H.-K. Chung, K. Kirby, and J. F. Babb}
\address{
Institute for Theoretical Atomic and Molecular Physics,\\
Harvard-Smithsonian Center for Astrophysics,\\ 
60 Garden Street, Cambridge, MA 02138}

\maketitle
%
\begin{abstract}
Absorption of radiation from the sodium dimer molecular states
correlating to Na(3s)-Na(3s) is investigated theoretically.
Vibrational bound and continuum transitions 
from the singlet $\Xstate$  state 
to the first excited $\Astate$ and $\Bstate$  states
and from the triplet $\astate$ state 
to the first excited $\bstate$ and $\cstate$ states 
are studied quantum-mechanically.
Theoretical and experimental data are used to characterize 
the molecular properties taking advantage of knowledge 
recently obtained from {\em ab initio\/} calculations,
spectroscopy, and ultra-cold atom collision studies.
The quantum-mechanical calculations are carried out 
for temperatures in the range from 500 to 3000~K and are
compared with previous calculations and measurements where available.
\end{abstract}

\narrowtext
%
%
\section{INTRODUCTION}

Vast amounts of experimental 
spectroscopic data on the electronic states and
ro-vibrational levels of the sodium dimer are available and many
theoretical studies have been performed.  For example,
Ref.~\cite{VerBahRaj83} presents an extensive bibliography summarizing
a variety of work dating from 1874 to 1983. 
Nevertheless, recent developments in atom trapping and 
cold atom spectroscopy have led to 
improved atomic and molecular data
through combinations of cold collision data, photoassociation
spectroscopy, and magnetic field induced Feshbach resonance data
\cite{KhaBabDal97,Gut99,MoeVer95,CotDal94,TieWilJul96,VanAbeVer99,CruDulMas99}.

Now that very reliable information is available, 
calculations of absorption spectra at high temperatures
become feasible.
Absorption coefficients in absolute units 
for a gas of sodium atoms and molecules at temperatures
from 1070 to 1470~K were measured over the range of wavelengths from
350 to 1075~nm by Schlejen {\em et al.\/}~\cite{SchJalKor87}.  
They performed semiclassical calculations involving the relevant 
molecular singlet and triplet transitions, however, 
those previous calculations do not fully reproduce
their experimental spectra~\cite{SchJalKor87}.  
The present work is concerned with the absorption 
involving two ground Na (3s) atoms and 
a ground Na (3s) atom and an excited Na (3p) atom, 
corresponding to transitions between
the singlet transitions from the $\Xstate$ state to the $\Astate$ and
$\Bstate$ states and the triplet transitions from the $\astate$ to the
$\bstate$ and $\cstate$ states. We assembled and evaluated the
available data for the molecular system and calculated
quantum-mechanically the absorption spectra at temperatures between
500 and 3000~K.

%
%
\section{ABSORPTION COEFFICIENTS}

The thermally averaged absorption coefficients $k_{\nu}$
for molecular spectra at wavelength $\nu$
are obtained from the product of the thermally averaged cross sections 
and the molecular density~\cite{ChuKirBab99}.
In turn, the molecular density can be expressed 
in terms of the atomic density squared 
and the chemical equilibrium constant\cite{Gar95}.
In the present study, we use the atomic density-independent 
reduced absorption coefficient, $\case{k_{\nu}}{n_a^2}$,
where ${n_a}$ is atomic density.

Four possible types of vibrational transitions between 
two electronic states can be identified:
bound-bound (bb), bound-free (bf), free-bound (fb) and free-free (ff)
and quantum-mechanical expressions for 
the reduced absorption coefficient can be derived.
The radial wave function $\phi$ for a bound level,
with vibrational and rotational quantum numbers, $v$ and $J$, 
is obtained from the
Schr\"{o}dinger equation for the relative motion of the nuclei,
\begin{equation}
\label{e:schrodinger}
\frac{d^2 \phi_{vJ\Lambda}(R)}{dR^2} + 
\left(2\mu E_{vJ\Lambda} - 2\mu V(R) - \frac{J(J+1)-\Lambda^2}{R^2}
\right)\phi_{vJ\Lambda}(R) = 0, 
\end{equation}
where $V(R)$ is the potential for the relevant electronic state
labeled by the projection $\Lambda$ of the electronic orbital 
angular momentum on the internuclear axis,
$\mu$  is the reduced mass of the nuclei,
and $E$ is the eigenvalue of the bound level or the continuum energy.

For the temperatures of interest here, $T \le  3000$~K,
the bound-bound reduced absorption coefficient 
from a vibration-rotation state 
of the lower electronic state $(v'',J'',\Lambda'')$ 
to the vibration-rotation state of 
the upper electronic state $(v',J',\Lambda')$ is 
\cite{ChuKirBab99,SanDal71,Doy68,LamGalHes77}
\begin{eqnarray} \label{e:bound}
\frac{k_{\nu}^{bb}}{n_a^2} & = &\frac{C(\nu)}{h} f(k_BT) \exp(D_e/k_BT)\nonumber \\
&\times& \sum_{v''~J''}\sum_{v'}   \omega_{J''}(2J''+1)\exp(-E_{v''J''}/k_BT)
  \left|\langle\phi_{v''J''\Lambda''}\right|D(R)
  \left|\phi_{v'J'\Lambda'} \rangle\right|^2 g(\nu-\bar{\nu}) ,
\end{eqnarray}
where 
$\nu$ is the frequency, $\bar{\nu}$ is the 
transition energy of the bound-bound transition, 
\begin{equation}
C(\nu)= \frac{(2-\delta_{0,\Lambda' + \Lambda''})}{2-\delta_{0,\Lambda''}}
	\frac{8\pi^3\nu}{3c}
\end{equation}
and ~\cite{MieJul82}
\begin{equation}
f(k_BT) =\frac{(2S_{m}+1)}{(2S_{a}+1)^2}
	\left[\frac{h^2}{2\pi\mu k_BT}\right]^{3/2},
\end{equation}
$S_m$ and $S_a$ are spin multiplicities for, 
respectively, the Na molecule and the Na atom,
and $k_B$ is Boltzmann constant.
The $Q$-branch approximation ($J' = J''$) is used 
and the line-shape function $g(\nu-\bar{\nu})$ 
is replaced by $1/\Delta\nu$.
In evaluating Eq.~(\ref{e:bound}) at some $\nu_i$ 
on the discretized frequency interval,
all transitions within the frequency range $\nu_i-\frac{1}{2}\Delta\nu$ 
to $\nu_i+\frac{1}{2}\Delta\nu$ are summed 
to give a value $\frac{k^{bb}(\nu_i)}{n_a^2}$.
The nuclear spin statistical factor 
$\omega_{J}$ for ${}^7\Namol$ with $I=\case{3}{2}$ is 
$[I/ (2I+1)]=\case{3}{8}$ for even $J$ and 
$[(I+1)/ (2I+1)]=\case{5}{8}$ for odd  $J$. 

The bound-free absorption coefficient 
from a bound level of the lower electronic state $(v''J''\Lambda'')$ 
to a continuum level of the upper electronic state $(\epsilon'J'\Lambda')$ 
is
\begin{eqnarray}
\frac{k_{\nu}^{bf}}{n_a^2} &=&C(\nu) f(k_BT)  \exp(D_e/k_BT) \nonumber \\
&\times& \sum_{v''~J''}  \omega_{J''}(2J''+1)\exp(-E_{v''J''}/k_BT)
  \left|\langle\phi_{v''J''\Lambda''}\right|D(R)
  \left|\phi_{\epsilon'J'\Lambda'} \rangle\right|^2.
\end{eqnarray}
The continuum wave function is energy normalized. 
For a free-bound transition and free-free transition, respectively,
\begin{eqnarray}\label{e:free-bound}
\frac{k_{\nu}^{fb}}{n_a^2} &=& C(\nu) f(k_BT) \nonumber \\
&\times&  \sum_{v'~J'} \omega_{J''}(2J''+1)\exp(-\epsilon''/k_BT)
  \left|\langle\phi_{\epsilon''J''\Lambda''}\right|D(R)
  \left|\phi_{v'J'\Lambda'} \rangle\right|^2
\end{eqnarray}
and
\begin{eqnarray}\label{e:free}
\frac{k_{\nu}^{ff}}{n_a^2} &=& C(\nu) f(k_BT) \nonumber \\
&\times& \sum_{J'} \int d\epsilon' \omega_{J''}(2J''+1)\exp(-\epsilon''/k_BT)
  \left|\langle\phi_{\epsilon''J''\Lambda''}\right|D(R)
  \left|\phi_{\epsilon'J'\Lambda'} \rangle\right|^2.
\end{eqnarray}
%
%
\section{Molecular Data}

\begin{figure}
\centering
\epsfxsize=.85\textwidth\epsfbox{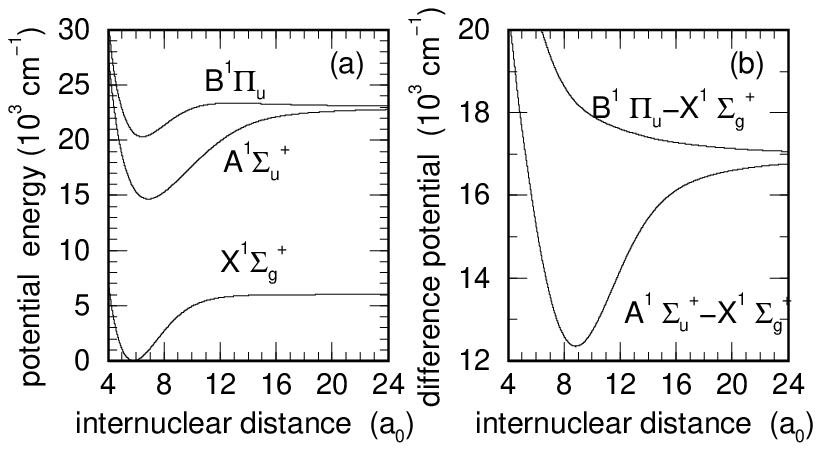}
\caption{(a) Adopted potentials $V(R)$ for the $\mbox{X}\,{}^1\Sigma_g^+$,
$\mbox{A}\,{}^1\Sigma_u^+$, and $\mbox{B}\,{}^1\Pi_u$ electronic
states.
(b) Difference potentials $V_{\Bstate}(R)-V_{\Xstate}(R)$
and $V_{\Astate}(R)-V_{\Xstate}(R)$.\label{sing-pots}}
\end{figure}

\begin{figure}
\centering
\epsfxsize=.85\textwidth\epsfbox{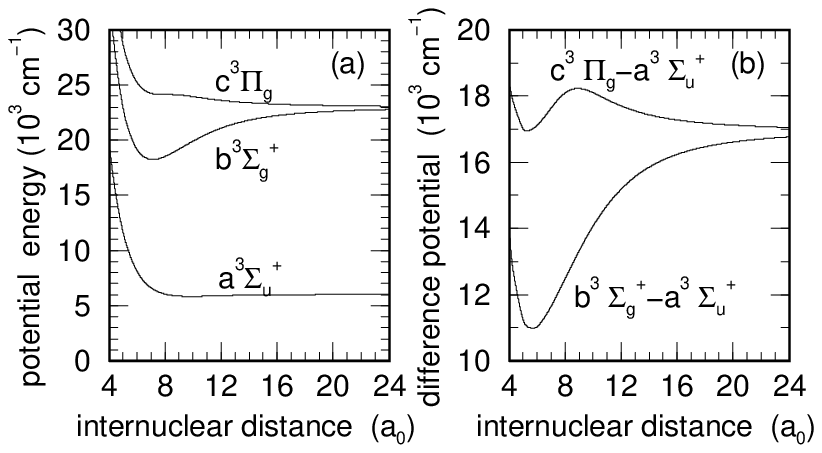}
\caption{
(a) Adopted potentials $V(R)$ for the $\astate$, $\bstate$ and
$\cstate$ states.
(b) Difference potentials $V_{\bstate}(R)-V_{\astate}(R)$
and $V_{\cstate}(R)-V_{\astate}(R)$.\label{trip-pots}}
\end{figure}

The adopted singlet $\Xstate$, $\Astate$, and $\Bstate$ potentials
and the differences of the upper potential and the lower $\Xstate$ potential
(difference potentials or transition energies) 
are plotted in Fig.~\ref{sing-pots}.
The adopted triplet $\astate$, $\bstate$, and $\cstate$ potentials
and the difference potentials are plotted in Fig.~\ref{trip-pots}.
In the remainder of the section 
details on the construction of the potentials are given.
We use atomic units throughout.

\subsection{The $\Xstate$ potential}

For $R < 4~a_0$, we adopted a short range form  $a\exp (-bR)$, 
with $a = 2\,702\,514.0 ~\invcm$ and $b = 2.797\,131$~\AA$^{-1}$ 
as given by Zemke and Stwalley~\cite{ZemStw94}.
Over the range of $4<R<30~a_0$ we used 
the Inverse Perturbation Analysis (IPA) potential given by
van Abeelen and Verhaar\cite{VanAbeVer99} 
which is consistent with data from photoassociation
spectroscopy, molecular spectroscopy, and magnetic-field induced
Feshbach resonances in ultra-cold atom collisions. 
For the long range form, we used 
\begin{equation}\label{e:long-range}
-C_6/R^6 - C_8/R^8 - C_{10}/R^{10} - AR^{\frac{7}{2}\alpha -1}\exp(-2\alpha R),
\end{equation}
where $C_6=1\,561$~\cite{KhaBabDal97}, 
$C_8=111\,877$, $C_{10} = 11\,065\,000$~\cite{MarSadDal94},  
$A = \case{1}{80}$, and  $\alpha=0.626$ 
\cite{VanAbeVer99,SmiChi65}.
To fit the very accurate dissociation energy, $6\,022.0286(53) ~\invcm$, 
recently measured by Jones {\em et al.\/}~\cite{JonMalBiz96},
a point at the potential minimum $5.819\,460 ~a_0$ was added.
The short and long-range data were smoothly connected to the IPA values.
Vibrational eigenvalues calculated with our adopted potential
agree for $v \le 44$ to within 0.1 $\invcm$ with 
published Rydberg-Klein-Rees (RKR) values~\cite{ZemStw94}.
Our final potential yielded an $s$-wave scattering length of 15~$a_0$ 
in satisfactory agreement with the accepted value of 
$19.1\pm 2.1~a_0$~\cite{VanAbeVer99}.

\subsection{The $\Astate$ potential}

We used {\em ab initio\/} calculations given by
Konowalow {\em et al.\/}~\cite{KonRosHoc80}
for values of $R$ over the range $3.8~a_0 <R< 4.75~a_0$.
We combined the RKR potential values 
over the range $2.522\,19~$\AA $<R<7.204\,14~$\AA
~given by Gerber and M{\"o}ller~\cite{GerMol85} with
the RKR potential values over the range 
of $7.260\,536~$\AA $<R<261.327\,403~$\AA
~given by Tiemann, Kn{\"o}ckel, and Richling~\cite{TieKnoRic96,Tie98}.
The data was connected to the long range form,
\begin{equation}
-C_3/R^3 -C_6/R^6 - C_8/R^8, 
\end{equation}
with the values of $C_3=12.26$, $C_6=4\,094$ and $C_8=702\,500$
\cite{MarDal95}.
For $R< 3.8~a_0$, the form $a\exp(-bR)+c$ was used with 
the parameters $a = 0.9532$, $b=0.5061$ and $c=0.104696$
computed to smoothly connect to the RKR points.
The adopted potential yields a value of 
$D_e = 8\,297.5~\invcm$ using 
$T_e = 14\,680.682~\invcm$~\cite{GerMol85} 
and the atomic asymptotic energy of $16\,956.172~\invcm$~\cite{NIST}.
The  calculated eigenvalues reproduce the input RKR values to
within 0.4~$\invcm$. 
For the transition frequencies measured by Verma, Vu, and Stwalley
\cite{VerVuStw81} and by Verma {\em et al.\/}~\cite{VerBahRaj83}
over a range of vibrational bands we find typical agreement to 
about 0.4~$\invcm$ for $J'$ values up to 50 
increasing to 1~$\invcm$ for $J'=87$.
We also have good agreement with less accurate measurements
by Itoh {\em et al.\/}~\cite{ItoFukUch83}.
One precise transition energy measurement is available:
In a determination of the dissociation energy of the sodium molecule
Jones {\em et al.\/}~\cite{JonMalBiz96} measured the value
18762.3902(30)~$\invcm$ for the $v'=165, J'=1$ to
$v''=31,J''=0$ transition energy. Our value of 
18762.372~$\invcm$ is in excellent agreement.

\subsection{The $\Bstate$ potential} 

The RKR potential of Kusch and Hessel~\cite{KusHes78}
was used\footnote{
For this reference, we correct an apparent typographical error 
of $4.309\,78$~\AA~with $4.339\,78$~\AA ~obtained by comparison 
with RKR potential of Demtr{\"o}der and Stock~\cite{DemSto75}.}
over the range of $2.655\,5671~\mbox{\AA{}}<R<5.173\,513\,4~$\AA.
For the values of $R$ in the ranges 
$2.581~\mbox{\AA{}}<R<2.646\,0268~$\AA ~and 
$5.251\,918\,4~\mbox{\AA{}}<R<11.0~$\AA, 
we took the potential values from Tiemann~\cite{Tie87}.
We also took his long-range form,
\begin{equation}
C_3/R^3 -C_6/R^6 + C_8/R^8 - a\exp(-bR),
\end{equation}
with  $C_3 = 6.1486$, $C_6=6490.5$, $C_8=868135.2$, 
$a=23.7011$, and $b=0.7885$.
For $R < ~2.581~$\AA, the form  $a \exp(-bR) + c$ was used with 
the values $a = 14.97332$, $b=1.42983$  and $c =0.0121935$ 
chosen to give a smooth connection with the data from Tiemann.

The $\Bstate$ potential exhibits a barrier that has been studied 
extensively~\cite{GerMol85,DemSto75,Tie87,KelWei84} and
the maximum value occurs around $R=13~a_0$ (6.9~\AA)
as shown in Fig.~\ref{sing-pots}.
We took $D_e = 2\,676.16~\invcm$ using $T_e = 20\,319.19~\invcm$ 
from Kusch and Hessel~\cite{KusHes78} and
the barrier energy $371.93~\invcm$ measured from dissociation  
given by Tiemann~\cite{Tie87}.
The calculated energy $23\,393.524~\invcm$ of the $v'=31$, $J'=42$ state 
with respect to the $\Xstate$ state potential minimum 
compares well to the measured value,  $23\,393.650 ~\invcm$.
Quasibound levels from $v'=24$ to $v'=33$ for  the several $J'$ values 
observed by Vedder {\em et al.}\cite{VedChaFie84} 
are reproduced to within $0.1~\invcm$ and 
calculated transition frequencies compare well, 
to within  $0.5~\invcm$, with those measured by 
Camacho {\em et al.\/}~\cite{CamPoyPol96}.

\subsection{The $\astate$ potential} 

RKR potentials are available from Li, Rice, and
Field~\cite{LiRicFie85} and Friedman-Hill and Field~\cite{FriHilFie92}
and a hybrid potential was constructed by Zemke and
Stwalley~\cite{ZemStw94} using various available data.
An accurate {\em ab initio\/} study was carried out by
Gutowski~\cite{Gut99} for $R$ values between 2 and 12.1~\AA~ 
and the resulting potential has well depth 
$176.173~\invcm$ and equilibrium distance 5.204~\AA.

Our adopted potential consists of Gutowski's 
potential connected to the long-range form given in 
Eq.~(\ref{e:long-range}) with the values 
for $C_6$, $C_8$, $C_{10}$ and $\alpha$ the same 
as for the $\Xstate$ state, but with $A=-\case{1}{80}$.
For $R<2$~\AA~ the short range form $a\exp(-bR)$ was used
with $a = 1.4956$ and $b = 0.79438$ chosen to smoothly 
connect to the {\it ab initio} data.
Our adopted potential yields an s-wave scattering length of 
65~$a_0$ in agreement with the value $65.3\pm 0.9$ of
van Abeleen and Verhaar~\cite{VanAbeVer99}.
Recently, a potential alternative to Gutowski's
was presented by Ho {\em et al.\/}~\cite{HoRabSco00}.
For the present study the two potentials are comparable---we 
will explore their differences in a subsequent publication.

\subsection{The $\bstate$ and $\cstate$ potentials} 
 
We are unaware of empirical excited state triplet potentials but
{\it ab initio} calculations are available 
from Magnier {\em et al.\/}~\cite{MagMilDulMas93},
Jeung~\cite{Jeu83} and Konowalow {\em et al.\/}~\cite{KonRosHoc80}.
Comparing the available potentials, 
we found for the $\bstate$ state that 
the experimentally~\cite{FarDem97} determined 
$T_e$ of $18\,240.5~\invcm$ and $D_e$ of $4\,755~\invcm$
are closest to Magnier's calculated values 
($T_e$ of $18\,117~\invcm$ and $D_e$ of $4\,740.7~\invcm$) 
compared with Jeung's 
($T_e$ of $18\,400~\invcm$ and $D_e$ of $4\,702.4~\invcm$) and 
Konowalow {\em et al\/.}'s
($D_e$ of $4\,599~\invcm$).
Also, we found Magnier's potential gave the best agreement 
with experimental measurements~\cite{FarKocPla94} of the 
term differences of the $\astate(v'') \rightarrow \bstate(v')$ 
vibrational transitions.
For the $\bstate$ state and, in the absence of experimental data
for the $\cstate$ potential, we adopted Magnier's calculated potentials
over the range of $R$ values $5 <R< 52~a_0$.
Over the range of $R$ values $4.25<R< 5~a_0$,
we used potentials by  Konowalow {\em et al.\/}~\cite{KonRosHoc80}.
For the $\bstate$ and $\cstate$ adopted potentials,
the long-range form was taken from Marinescu and Dalgarno~\cite{MarDal95}
for $R > 52~a_0$ and for $R < 4.25~a_0$, 
we used the form $a\exp(-bR)$ where 
the values are $a = 55.7864$ and $b = 1.75934$ for the $\bstate$ potential 
and $a = 2.67691$ and $b = 0.91547$ for the $\cstate$ potential.
%
%
\subsection{Transition dipole moment functions}

We used for the singlet transitions the {\em ab initio\/} calculations of
Stevens {\em et al.\/}~\cite{SteHesBer77} over the range $2<R<12$~$a_0$.
For $R>12$ the transition dipole moment functions were
approximated by $a + b/r^3$,  where $a = 3.586\,4$ and $b = 284.26$ 
for $\Xstate \rightarrow \Astate$ transitions
and $a = 3.5017$ and $b = -142.13$ 
for $\Xstate \rightarrow \Bstate$ transitions. 
The parameter values  for $a$ were selected to match the short-range parts 
and those for $b$ were from Marinescu and Dalgarno~\cite{MarDal95}.
The $\Xstate \rightarrow \Astate$ dipole moment function was scaled with a
factor of 1.008, as discussed in Sec.~\ref{lifetimes} below.
For the triplet transitions the {\em ab initio\/} calculations of 
Konowalow {\em et al.\/}~\cite{KonRosHoc83} were used 
over the range $4<R<100$~$a_0$.
%
%
\section{Results}

%
\subsection{Lifetimes}\label{lifetimes}

In order to evaluate our assembled potential energy and transition
dipole moment data we calculated lifetimes of ro-vibrational levels of
the $\Astate$ and $\Bstate$ states and compare with prior studies. 

\begin{figure}
\centering
\epsfxsize=.85\textwidth\epsfbox{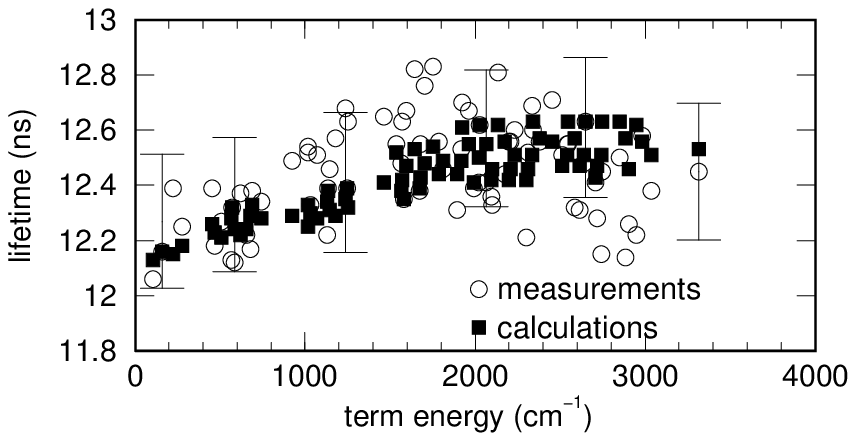}
\caption{
Comparisons of calculated lifetimes of $\Astate$ ro-vibrational levels 
with experimental measurements~\protect\cite{BauKorPre84}.\label{A-lifetimes}}
\end{figure}

Lifetimes for levels of the $\Astate$ state have been
measured~\cite{VerVuStw81,DucLitZim76,Woe78,BauKorPre84} and
calculated~\cite{VerVuStw81,Par99}.
In Fig.~\ref{A-lifetimes} we present a comparison of rotationally
resolved lifetimes for levels of the $\Astate$ state measured by
Baumgartner {\em et al.\/}~\cite{BauKorPre84} with the present
calculations.  
In evaluating the lifetimes, we used the procedures described in 
Ref.~\cite{ChuKirBab99}.
When the transition dipole moment function of Ref.~\cite{SteHesBer77}
is multiplied by a factor of 1.008,  
agreement is generally very good over the range $0$ to $3500~\invcm$ 
of available term energies. 
Our calculations are also in good agreement with the rotationally 
unresolved measurement of Ducas {\em et al.\/}~\cite{DucLitZim76}
and the calculations using different molecular data by Pardo~\cite{Par99}.

\begin{figure}
\centering
\epsfxsize=.85\textwidth\epsfbox{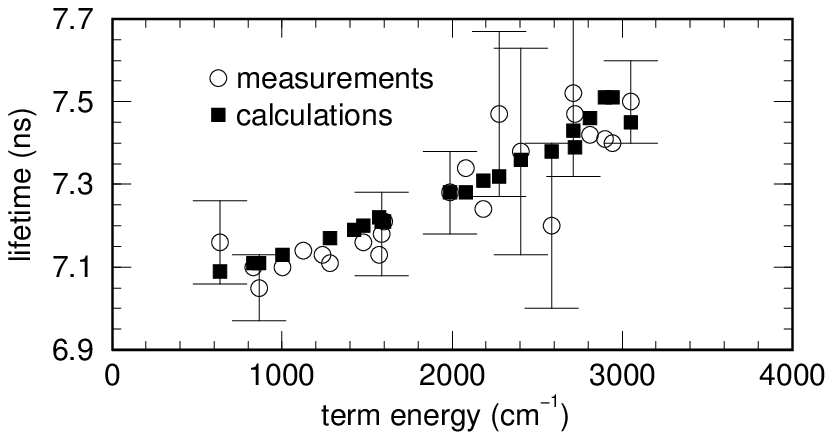}
\caption{
Comparisons of calculated lifetimes of $\Bstate$ ro-vibrational levels 
with experimental measurements~\protect\cite{DemSteSto76}.\label{B-lifetimes}}
\end{figure}

Rotationally resolved lifetimes for the $\Bstate$ state 
have been measured by Demtr{\"o}der {\em et al.\/}~\cite{DemSteSto76}.
Demtr{\"o}der {\em et al.\/} found that 
the lifetimes for the $\Bstate$ state are sensitive to 
the slope of the transition dipole moment function 
in the range of internuclear separation from, 
roughly, $4 <R< 10~a_0$ and 
they obtained an empirical value for the function
that we found to be in good agreement with 
the transition dipole moment function of 
Stevens {\em et al.\/}~\cite{SteHesBer77}.
Using the transition dipole moment function of 
Stevens {\em et al.\/}, in turn, 
we find good agreement between our calculated lifetimes
and experimental lifetime measurements~\cite{DemSteSto76}, 
as shown in  Fig.~\ref{B-lifetimes}. 
The {\em ab initio\/} dipole moment of 
Konowalow {\em et al.\/}~\cite{KonRosHoc83} was found not to 
reproduce the experimental lifetimes.
%
%
\subsection{Absorption Coefficients}

Absorption spectra in the far-line wings of the Na(3s)-Na(3p)
resonance lines are investigated in terms of singlet
and triplet molecular transitions.
The blue wing  consists of radiation from 
$\Xstate\rightarrow\Bstate$ and $\astate\rightarrow\cstate$ transitions and
the red wing  from 
$\Xstate\rightarrow\Astate$ and $\astate\rightarrow\bstate$ transitions.
There are few experimental studies
~\cite{SchJalKor87,LamGalHes77,WoedeG81}
and that of Schlejen {\em et al.\/}~\cite{SchJalKor87}
is most relevant to our work.
In this section, we compare our calculated 
absorption coefficients with the measurements of
Schlejen {\em et al.\/}~\cite{SchJalKor87}.

\begin{figure}
\centering
\epsfxsize=.8\textwidth\epsfbox{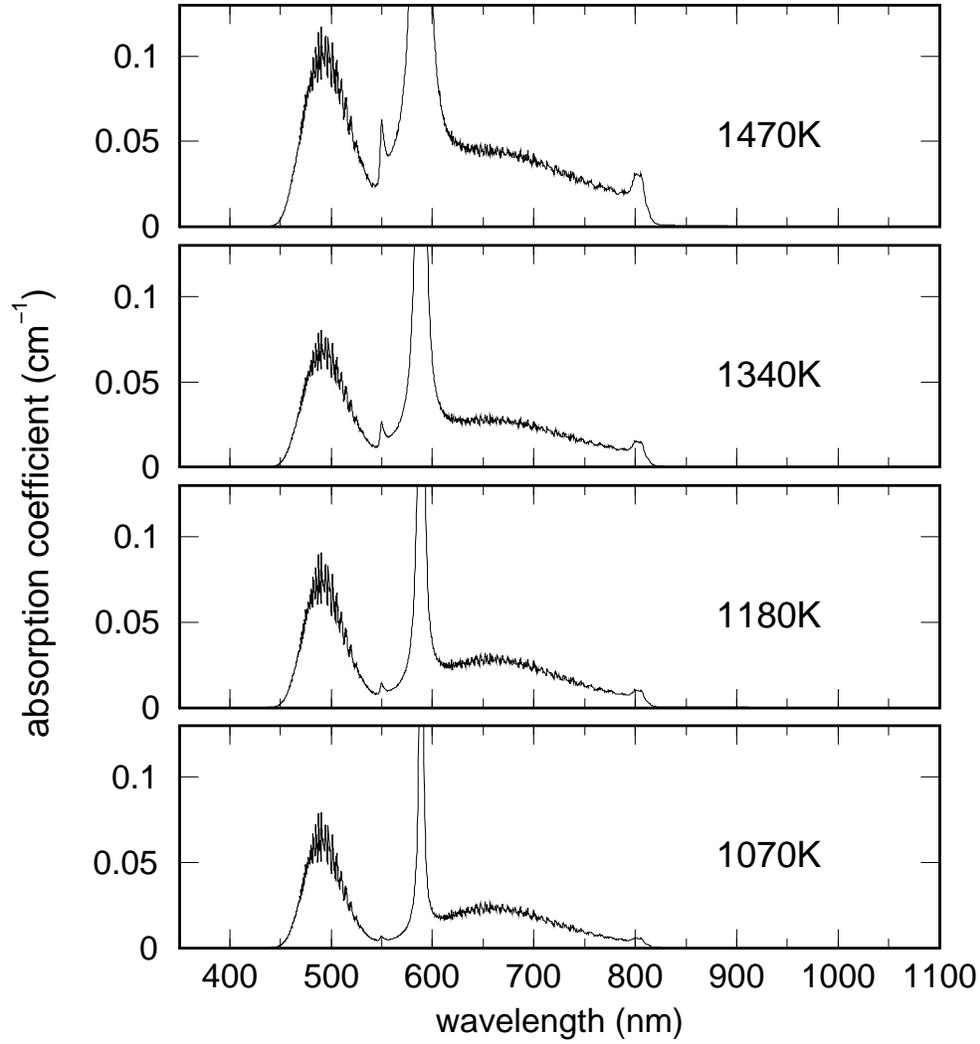}
\caption{
Absolute values of the absorption coefficient 
are shown for four different temperatures 
for a comparison with experimental spectra 
reported in Figure. 5 of Ref.~\protect\cite{SchJalKor87}.
The calculations were performed with bin size $\Delta\nu = 10~\invcm$. 
\label{schlejen}}  
\end{figure}

The theoretical spectra are assembled from 
four molecular band spectra over the wavelength range 
450--1000~nm excluding the region 
589$\pm$2~nm around the atomic resonance lines.
The far line wings are calculated using 
Eqs.~(\ref{e:bound})--(\ref{e:free}), with the data from Sec.~III.
In the calculations, all the vibrational levels including 
quasi-bound levels with rotational quantum numbers up to $250$ are included. 
The maximum internuclear distance that is used for integration of 
the transition dipole matrix element is approximately $100~a_0$ and
the Numerov integration used to obtain the energy-normalized 
continuum wave function is carried out to  $100~a_0$ at which
the wave function is matched to its asymptotic form.
The bin size $\Delta\nu$ used for Eq.~(\ref{e:bound})
was 10~$\invcm$ simulating the experimental resolution.
Results for absolute absorption coefficients
computed with the quoted atomic densities and 
temperatures of Schlejen {\em et al\/}~\cite{SchJalKor87} 
and shown in Fig.~\ref{schlejen} compare very well
with the four experimental spectra given by Schlejen {\em et al\/},
given in Figure. 5 of Ref.\cite{SchJalKor87}. 
The spectra show clearly that as temperature increases, certain 
satellite features grow more apparent at 551.5~nm and 804~nm.
These satellites will be discussed in greater detail later 
in this section.
The present calculations reproduce fine-scale ro-vibrational
features present but unresolved in the measurements of Ref.\cite{SchJalKor87}.

\begin{figure}
\centering
\epsfxsize=.99\textwidth\epsfbox{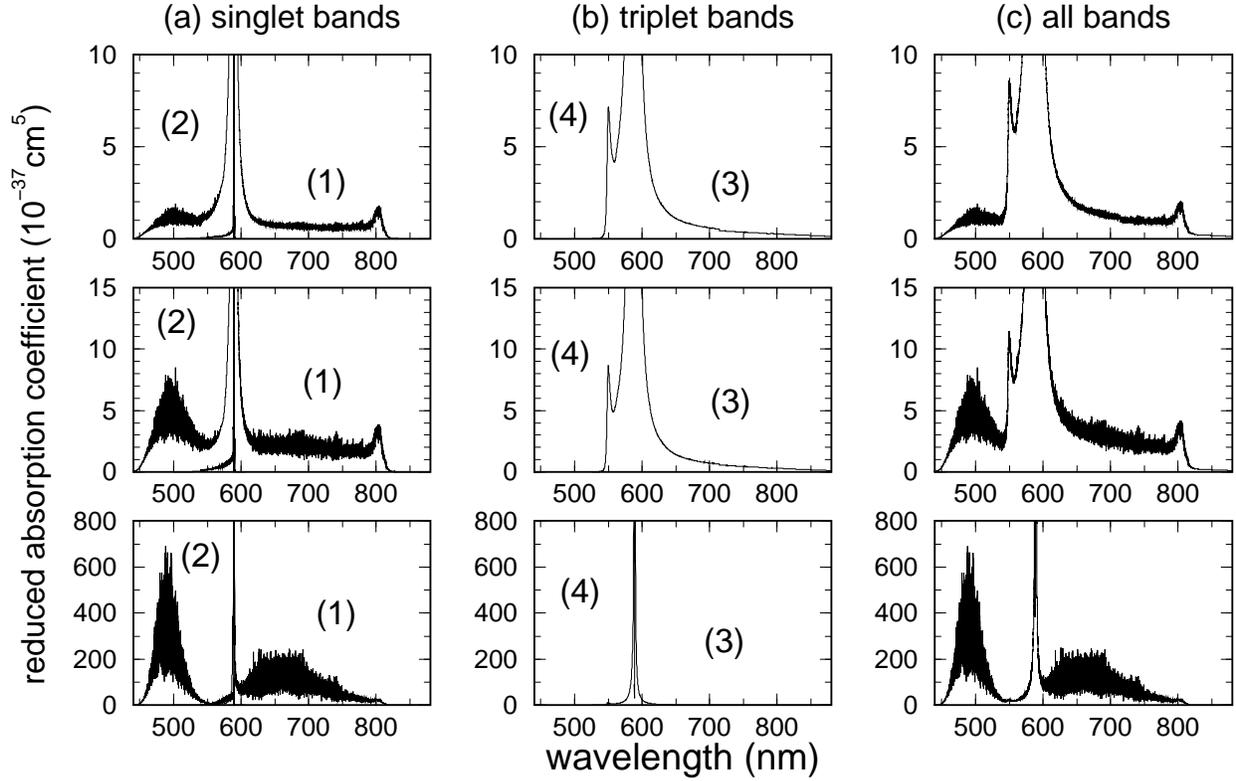}
\caption{
Contributions to the reduced absorption coefficient 
at 1000~K (bottom plots), 2000~K (center plots) and 
3000~K (top plots) from molecular band radiation from 
(a) the singlet bands, 
$\Xstate\rightarrow\Astate$(1) and $\Xstate\rightarrow\Bstate$(2)
transitions, and 
(b) the triplet bands,
 $\astate\rightarrow\bstate$(3) and $\astate\rightarrow\cstate$(4) 
transitions. 
The total of the singlet and triplet bands is shown in (c).
Note that the scale for the reduced absorption coefficient at 1000~K is
very much greater than the scale at 2000~K and 3000~K.
The calculations were performed with bin size $\Delta\nu = 1~\invcm$. 
\label{mol-bands}}
\end{figure}

We also have calculated reduced absorption coefficients at temperatures 
up to 3000 K using the bin size $\Delta\nu$ of 1~$\invcm$
for Eq.~(\ref{e:bound}).
The contributions of the four molecular bands to the reduced absorption 
coefficients are shown in Fig.~\ref{mol-bands} 
for three temperatures $1000$~K, $2000$~K and $3000$~K.
As can be seen by comparing columns (a) and (b) 
in Fig.~\ref{mol-bands},
the singlet transitions contribute more 
to the reduced absorption coefficients in the far line wings
and the triplet transitions contribute more
near the atomic resonance lines.
We found that for singlet transitions 
bound-bound and bound-free transitions
are dominant over free-bound and free-free transitions
for the temperature range $T \le 3000$~K,
thus accounting for the ``grassy'' structure 
in Fig.~\ref{mol-bands}(a).
However, the free-bound and free-free contributions 
increase rapidly with temperature.
In contrast to the singlet transitions, 
the triplet transitions arise mainly from 
free-bound and free-free transitions
due to the shallow well of the initial $\astate$ state.
Hence, the reduced absorption coefficients
in Fig.~\ref{mol-bands}(b) do not exhibit much structure.
Because the density of bound molecules decreases rapidly
with increasing temperature, 
the reduced absorption coefficient in the line wings
due to the singlet transitions also decreases rapidly
with increasing temperature.
It should be noted that the scale of the reduced absorption
coefficient at 1000~K is two orders of magnitude larger
than the scale shown for $T$ = 2000~K and 3000~K.

Woerdman and De Groot~\cite{WoedeG81} 
derived the reduced absorption coefficient at 2000~K 
from a discharge spectra. 
The measured values of $5 \pm 1 \times 10^{-37}~\invcm$ 
at $500$~nm and $10 \pm 1 \times 10^{-37}~\invcm$ 
at $551.5$~nm are well-reproduced by our values of,
respectively, $5 \times 10^{-37}~\invcm$ 
and $11 \times 10^{-37}~\invcm$ 
calculated with the bin size $\Delta\nu$ of 5~$\invcm$  
simulating the experimental resolution 
obtained by Woerdman and De Groot~\cite{WoedeG81}. 

The molecular absorption spectra contain ``satellite''
features around the energies where the difference potentials
possess local extrema~\cite{SanWor73,SzuBay75}.
For Na$_2$ the satellite frequencies have been studied
~\cite{SchJalKor87,LamGalHes77,WoedeG81,DeGSchWoe87}
and the energies have been calculated using {\it ab initio}
methods~\cite{KonRosHoc83}.
In the present work, we investigate the satellites
arising from $\astate \rightarrow \cstate$, 
$\Xstate \rightarrow \Astate$, and $\astate \rightarrow \bstate$
transitions with measured maximum intensities at, respectively, 
the wavelengths $551.5$~nm, $804$~nm, and $880$~nm.
The calculated extrema of the difference potentials adopted in the
present study occur at wavelengths at 
$548$~nm, $809$~nm and $913$~nm, however
in the quantum-mechanical approach there is no well-defined singularity.

\begin{figure}
\centering
\epsfxsize=.85\textwidth\epsfbox{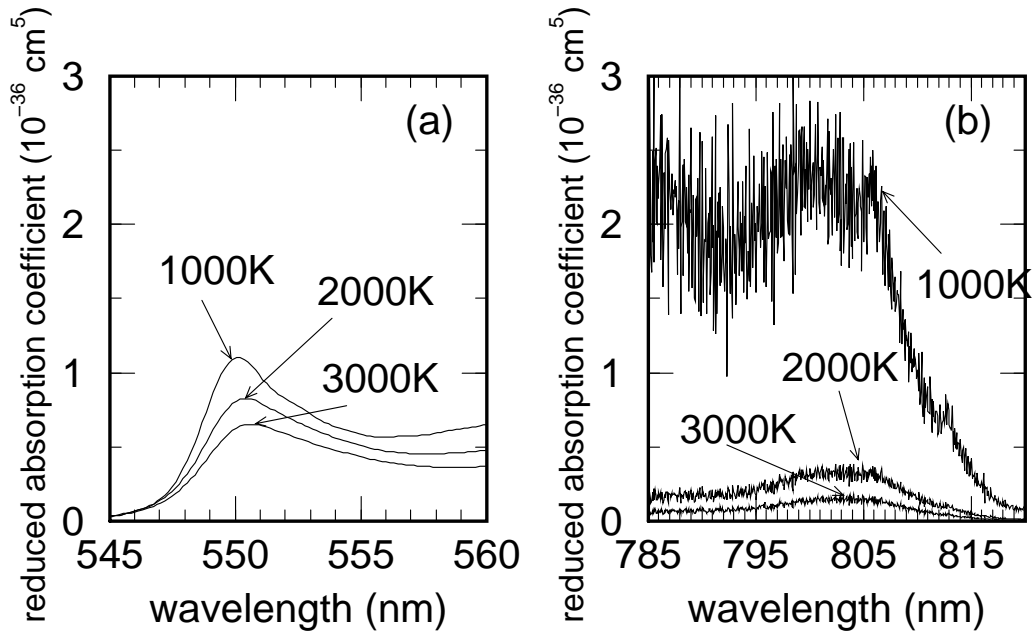}
\caption{(a) Calculated reduced absorption coefficients 
for the $\astate \rightarrow \cstate$ satellite for three temperatures.
(b) Calculated reduced absorption coefficients 
for the $\Xstate \rightarrow \Astate$ satellite for three temperatures. 
\label{satAX}}
\end{figure}

We can use the quantum-mechanical theory to study satellite features
in more detail and as a function of temperature.
In Fig.~\ref{satAX} we show calculated reduced absorption coefficients 
at three temperatures for the $\astate \rightarrow \cstate$ 
and $\Xstate \rightarrow \Astate$ transitions.
The rich ro-vibrational structure 
in the $\Xstate \rightarrow \Astate$ satellite feature
arises because the dominant contributions are from
bound-bound transitions; the structure is not reproduced by
semiclassical theories~\cite{SchJalKor87}.
In contrast, the smooth, structureless  
$\astate\rightarrow\cstate$ satellite feature is 
due mainly to free-free transitions, and 
consequently, the decrease of the satellite intensity with 
temperature is less severe.
The slight discrepancy between the calculated
wavelength of 550~nm and the measured wavelength of 
551.5~nm~\cite{WoedeG81,deGvVli75,VezRukMov80}
for the peak intensity is probably due to remaining uncertainties
in the triplet potentials~\cite{HoRabSco00,Tie00}.

\begin{figure}
\centering
\epsfxsize=.85\textwidth\epsfbox{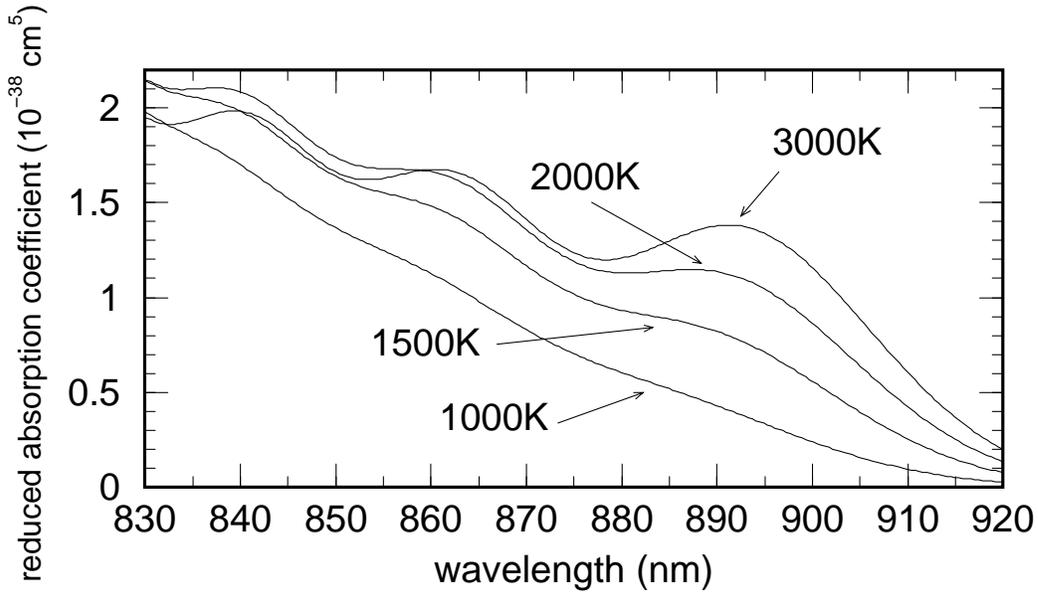}
\caption{
Reduced absorption coefficients near satellite structures from 
$\astate \rightarrow \bstate$ bands for four temperatures. 
Note that the scale is two orders of magnitude smaller than
in Fig.~7. \label{satba}}
\end{figure}

We also investigated the $\astate\rightarrow\bstate$ satellite
which is far weaker in intensity at $T \le 3000$~K than the 
$\Xstate\rightarrow\Astate$ and $\astate\rightarrow\cstate$ satellites.
The $\astate\rightarrow\bstate$ satellite 
arises primarily from free-bound transitions.
The population density of atom pairs with high continuum energies 
in the initial $\astate$ state increases with temperature, 
see Eq.~(\ref{e:free-bound}), and more ro-vibrational levels 
in the $\bstate$ state are accessible through absorption
of radiation, as can be seen from the potential curves 
shown in Fig.~\ref{trip-pots}(a).
As a result, this satellite feature exhibits an increase
in intensity with temperature. 
In Fig.~\ref{satba} calculated absorption coefficients
for temperatures 1000, 1500, 2000 and 3000~K are plotted.
The satellite feature intensity was measured at 1470~K
by Schlejen {\em et al.\/}~\cite{SchJalKor87}.
They observed a primary peak at 880~nm and a secondary peak at 850~nm, 
compared to our calculated values at 1500~K of 890~nm and 860~nm,
respectively. The 10~nm discrepancy in both peaks is probably 
a result of uncertainties in the short range parts of our adopted 
$\astate$ and $\bstate$ potentials.
Our calculations also demonstrate that the wavelengths of 
the peaks change with temperature, see Fig.~\ref{satba},
and that the primary peak from quantum-mechanical calculations
is less prominent than that obtained from semiclassical calculations 
exhibited in Figures 6(c) and 6(d) 
of Schlejen {\em et al.\/}~\cite{SchJalKor87}.
Our calculated reduced absorption coefficients appear to be 
in excellent agreement with the reduced absorption coefficients
interpolated from Figures 6(a) and 6(b) of 
Schlejen {\em et al.\/}~\cite{SchJalKor87}
using their quoted Na densities.
\section{Conclusions}

We have carried out quantum-mechanical calculations of the  
reduced absorption coefficients in sodium vapor at high temperatures.
Accurate molecular data are an important ingredient.
Comparisons with experiments~\cite{SchJalKor87,WoedeG81} 
are good, but the theory is not limited by the 
previous experimental resolution. 
Future work~\cite{BabShuChu00} will focus on comparisons of the present theory
and experiments currently on-going in our group~\cite{ShuParYos00}.
\acknowledgements 
We thank R.~C\^ot\'e and A. Dalgarno for helpful
communications and E. Tiemann for generously supplying us with
additional unpublished data.  
We are grateful to Dr. G. Lister, Dr. H. Adler, and Dr. W. Lapatovich 
of OSRAM SYLVANIA Inc. and Dr. M. Shurgalin, Dr. W. Parkinson,
and Dr. K. Yoshino for helpful discussions.
This work is supported in part by the
National Science Foundation under grant PHY97-24713 and by a grant to
the Institute for Theoretical Atomic and Molecular Physics at Harvard
College Observatory and the Smithsonian Astrophysical Observatory.

%
%
%

\clearpage
%
%
\clearpage
\clearpage
\clearpage
%
%
\clearpage

\end{document}